\documentclass[twocolumn,fleqn]{svjour3}    
\smartqed  
\usepackage{amsmath}
\usepackage{amssymb}
\usepackage{graphicx}
\usepackage{scrextend}
\newcommand{\half}{\frac{1}{2}}

\newcommand{\bc}{\begin{center}}
\newcommand{\ec}{\end{center}}
\newcommand{\bea}{\begin{eqnarray}}
\newcommand{\eea}{\end{eqnarray}}
\newcommand{\beas}{\begin{eqnarray*}}
\newcommand{\eeas}{\end{eqnarray*}}
\newcommand{\be}{\begin{equation}}
\newcommand{\ee}{\end{equation}}
\newcommand{\bseq}{\begin{subequations}}
\newcommand{\eseq}{\end{subequations}}
\newcommand{\bdm}{\begin{displaymath}}
\newcommand{\edm}{\end{displaymath}}
\newcommand{\ben}{\begin{enumerate}}
\newcommand{\een}{\end{enumerate}}
\newcommand{\bit}{\begin{itemize}}
\newcommand{\eit}{\end{itemize}}
\newcommand{\bds}{\begin{description}}
\newcommand{\eds}{\end{description}}
\newcommand{\bvb}{\begin{verbatim}}
\newcommand{\evb}{\end{verbatim}}
\newcommand{\BF}{\begin{figure}}
\newcommand{\EF}{\end{figure}}
\newcommand{\BTB}{\begin{table}}
\newcommand{\ETB}{\end{table}}

\newcommand{\BT}{\begin{theorem}}
\newcommand{\ET}{\end{theorem}}
\newcommand{\BA}{\begin{algorithm}\vspace{24pt}}
\newcommand{\EA}{\end{algorithm}}
\newcommand{\BL}{\begin{lemma}}
\newcommand{\EL}{\end{lemma}}
\newcommand{\BR}{\begin{remark}}
\newcommand{\ER}{\end{remark}}
\newcommand{\BC}{\begin{corollary}}
\newcommand{\EC}{\end{corollary}}
\newcommand{\BD}{\begin{definition}}
\newcommand{\ED}{\end{definition}}

\newcommand{\bee}{\begin{eqnarray}}
\newcommand{\eee}{\end{eqnarray}}

\newcommand{\LL}{\left(\frac{\Delta x}{2}\right)^2}

\newcommand{\I}{{\mathcal I}}

\newcommand{\Or}{{\mathcal O}}

\begin{document}

\title{Artificial Viscosity -- Then and Now
}
\subtitle{Modeling Nature's Shocks}


\author{   L.\,G. Margolin   \and N.\,M. Lloyd--Ronning
}


\institute{L.\,G. Margolin \at
              Los Alamos National Laboratory,
               Los Alamos, NM, 87545 \\
              Tel.: 505-665-1947\\
              \email{len@lanl.gov}           
           \and
          N.\,M. Lloyd-Ronning \at
               Los Alamos National Laboratory Los Alamos, NM, 87545\\
               University of New Mexico at Los Alamos\\
              Tel.: 505-667-1358\\
             \email{lloyd-ronning@lanl.gov}  
}

\date{Received: date / Accepted: date}

\maketitle

\begin{abstract}
In this paper, we recount the history of artificial viscosity, beginning with its origin
in previously unpublished and unavailable documents, continuing on to current
research and ending with recent work describing its physical basis that
suggests new directions for improvement.  We focus 
on the underlying ideas that recognize the finiteness of scale and of measurement.
\vspace{.12cm}

\noindent
{\bf Los Alamos report LA-UR-22-20270}

\keywords{Shocks \and Artificial Viscosity \and Finite Scale Theory}
\end{abstract}
\section*{Preface}\label{preface}
\begin{addmargin}[1em]{2em}
``We don’t see things as they are; we see them as we are." (Anonymous)
\end{addmargin}
\vspace{.12cm}

We celebrate the centennial anniversary of Richard Becker's seminal paper
on shock structure \cite{Becker}. Becker's paper is most often cited for his
clever solution of the Navier--Stokes equations for the planar shock.  However
Becker's main purpose was to cast doubt on the ability of the Navier--Stokes
equations to model shock structure.  Becker concluded from his solution that
a shock would be about one molecular mean free path wide, violating
the assumptions under which the Navier--Stokes equations are
derived from the Boltzmann equation \cite{Kremer}.
It would be more than 50 years before computer simulation \cite{Bird} and 
laboratory experiments  \cite{Alsmeyer,Schmidt} would show that physical 
shocks are measured to be twice the
width predicted by theory, validating Becker's 
assertion that something beyond the Navier--Stokes description is needed.

While the inaccuracy of Navier--Stokes as a model for shocks is generally accepted,
a more accurate continuum model is still not at hand.  Recently, the finite scale
equations \cite{observer} have achieved some success in predicting the width
and the shape of shocks as measured in the laboratory \cite{Young,Rankine}.
The finite scale theory is designed to be a differential model of the discrete
equations used in numerical simulation. One key development of the theory is 
the derivation of the artificial viscosity, showing that it has a physical origin in
addition to its numerical benefits \cite{reality}. 

The focus in this paper is on the ideas that underlie artificial viscosity, both as
a numerical strategy and as a physical phenomenon.
We describe those underlying ideas and provide copious references where 
mathematical details can be found. We build on Becker's original work that emphasizes 
the role of dissipation in accurately modeling finite shock structure, and offer the 
speculative conclusion that on length scales much larger than the molecular mean 
free path, it is in the discrete equations that an improved description of nature may 
be found.
\section{Introduction}\label{intro}
\begin{addmargin}[1em]{2em}
``What we observe is not nature itself, but nature exposed to our method of questioning."\\
(Werner Heisenberg \cite{Heisenberg})
\end{addmargin}
\vspace{.12cm}

Artificial viscosity is one of the oldest and most enduring concepts in computational
fluid dynamics (CFD). The earliest computer simulations of shocks by Los Alamos 
staff showed unphysical oscillations ({\em wiggles}) behind the shock front.  The first 
remedy was based on shock tracking, a cumbersome process that required continuous 
human interaction with the computer \cite{Morgan}.  Artificial dissipation as it was first 
termed  represented an effective automatic solution to eliminate the wiggles. 
In section~\ref{heroes} we will describe this early history. Some of this history is
newly recorded due to the recent release of formerly classified documents by
Los Alamos National Laboratory.  

The early simulations of shocks were done in Lagrangian {\em codes}\footnote{Numerical
programs for solving hydrodynamic problems are typically called codes.}
in which the computational mesh moves with the local fluid velocity and so adapts to
the large compressions within the shock profile.  Explicit artificial 
viscosity proved overly diffusive when implemented in Eulerian codes which operate on 
a fixed mesh.  The development of nonoscillatory methods for Eulerian codes in the 
early 1970s represented a revolution in CFD. In section~\ref{euler} we will detour
to discuss the ideas that underlie this seemingly independent solution to eliminating 
unphysical oscillations.

In section~\ref{ideas} we describe some current work designed to improve the
application of artificial viscosity.  These focus on two areas: 1) incorporating the ideas
of nonoscillatory differencing to Lagrangian codes to minimize the effect of the
artificial dissipation on smooth parts of the flow; 2) developing a tensor artificial
viscosity to better preserve the geometric symmetries of the flow.

The finite scale theory \cite {observer} was formulated to provide a theoretical basis 
for implicit large eddy simulation (ILES) \cite{book}, then a controversial methodology 
for simulating turbulent flows. However, the derivation of the finite scale equations
does not refer either to turbulence nor to numerical simulation.  In section~\ref{FV}
we review the derivation of the finite scale equations and its application to the
theory of shocks.  The finite scale equations provide new insights into the interpretation 
of artificial viscosity, which we detail in section~\ref{flux}, and its implementation
which we discuss in \ref{discuss}.  

We conclude the paper in section~\ref{conclude} with a brief summary and thoughts 
for future development.

\section{The Heroes of Artificial Viscosity} \label{heroes}
\begin{addmargin}[1em]{2em}
``Now it can be told"
(Gen. Leslie Groves, \cite{Groves})
\end{addmargin}
\vspace{.12cm}

\noindent
Here we describe the early contributions of five heroes of artificial viscosity,
Rudolf Peierls, Robert Richtmyer, John von Neumann, Rolf Landshoff and
William Noh.
\vspace{.2cm}

\noindent
{\bf Rudolf Peierls}: The need for artificial dissipation to smear shocks was recognized by 
von Neumann even before the ENIAC (the first electronic computer) was operational.  von 
Neumann had proposed several ideas to Peierls, who responded in a letter dated March,
1944. In the last paragraph of that letter, Peierls wrote:
\vspace{.12cm}

\begin{addmargin}[1em]{2em}
``Incidentally, have you thought at all about the following alternative way of avoiding 
discontinuity.  In actual fact, the shock front has a finite width because of the viscosity and 
thermal conductivity of the medium.  But artificially, assuming a viscosity very much larger 
than the actual, you can obtain instead of the discontinuity a front of a finite width."
\cite{Archer}
\end{addmargin}
\vspace{.12cm}

The idea that important features of a shock, in particular the shock velocity and the
Rankine--Hugoniot relations, (the jump conditions), are independent of the physical
viscosity had been known since the 1870 paper of Rankine \cite{Salas}. The
application of that independence has proved a fundamental insight for CFD.
\vspace{.12cm}

\noindent
{\bf Robert Richtmyer}: In the open literature of numerical shocks, the name Richtmyer
is invariably associated with von Neumann by virtue of their landmark paper of 1950
\cite{VNR}.  However, the release of several formerly classified Los Alamos reports
tells a more detailed story making it clear that, in the case of artificial viscosity, it
was Richtmyer alone who did the {\em heavy lifting}. In an effort to give proper
credit, we will discuss the reports written solely by Richtmyer \cite{LA671,LA699} 
separately here. The scope and breadth of Richtmyer's report is nicely captured in 
the abstract
to \cite{LA671}.    \vspace{.12cm}

\begin{addmargin}[1em]{2em}
``This report gives a method of handling the finite difference equations of hydrodynamics
of compressible fluids.  Shocks are automatically taken care of by the expedient of
introducing a (real or fictitious) dissipation term . . . The effect of this term is to ``smear"
the shock somewhat, but the Hugoniot--Rankine conditions are satisfied and the
entropy increase is correct;"
\cite{LA671}
\end{addmargin}
\vspace{.12cm}

\noindent
Among the insights of this first report, Richtmyer discusses the source of the observed 
post--shock oscillations, the importance of conservation and the associated issues
of numerical stability. 

Perhaps the most important (and unexpected) contribution is the form of the dissipative
term, which is quadratic in the velocity gradient as
contrasted with physical viscous dissipation which is linear.  Richtmyer {\em derives}
the quadratic form by requiring that shocks of varying strengths, i.e, Mach number,
should all have approximately the same width when measured in units of the grid
spacing $\Delta x$. Richtmyer's insight was driven by stability considerations;
however it is interesting that the width of physical shocks is also relatively
constant when measured in terms of the molecular mean free path
\cite{Schmidt,Alsmeyer}.  But Richtmyer could not have known that fact
as those measurements were first made more than 20 years after his report.

The first report \cite{LA671} was written in March of 1948 and quickly followed by
a second report \cite{LA699} in August of 1948.  The main purpose of that
report was to demonstrate that the artificial viscosity would be stable with
the proper choice of the dimensionless coefficient.  It is not explicitly stated
how the numerical calculations were performed, but it appears they were done
by hand rather than on an electronic computer.   Richtmyer notes that the
calculations were performed by Irene Stegun, who is better known as co--author
of the Handbook of Mathematical Functions (Abramowitz \& Stegun). In any
case, those calculations which comprised less than 20 computational cycles
were compelling.

Richtmyer gives explicit credit to von Neumann and Peierls for the idea of
shock capturing.  He also references Becker \cite{Becker} for the importance
of dissipation in producing a finite shock width.  Curiously, the possibility of
including an artificial thermal conduction is not mentioned even though the
simultaneous consideration of viscosity and thermal conduction is a
principal new result of Becker's derivation of shock profile.
\vspace{.12cm}

\noindent
{\bf John von Neumann}: The 1950 journal article by von Neumann and Richtmyer
can be seen to contain a subset of the results of Richtmyer's report \cite{LA671}.  In 
particular, the quadratic form of the artificial viscosity is not derived in the
paper, but rather is postulated and shown to satisfy certain numerical and
physical requirements.  Further, no simulations are presented to illustrate the 
effectiveness of artificial viscosity.  However the importance of this paper's
appearance in the open literature cannot be overstated.  The modern reader
may not appreciate the challenges of computing in this period, the lack of
capability and lack of access \cite{Goldstine}. We note that Becker's paper
\cite{Becker} is also cited here, though no reference is made to Peierl's insight.

In addition to the artificial dissipation and the linearized stability analysis, the
idea of the staggered mesh for Lagrangian simulations was openly introduced in
this landmark paper. 
\vspace{.12cm}

\noindent
{\bf Rolf Landshoff}: Landshoff's singular contribution to artificial viscosity
lies in an unpublished report \cite{Landshoff} in 1955 in which the modern
form of artificial viscosity, referred to as {\em linear plus quadratic Q} was
first formulated.

Landshoff is a curious choice for a hero of artificial viscosity.  He was neither
primarily a hydrodynamics nor a computational fluids physicist.  Further, the
theoretical development in his report is substantially incorrect.  Nevertheless, 
his intuition has proved correct and enduring \cite{Caramana}. Although the 
quadratic viscosity of Richtmyer controlled the unphysical post--shock 
oscillations, it did not totally eliminate them nor the troublesome overshoot 
that typically appeared at the head of the numerical shock.

Landshoff's idea was to add a second term to the artificial dissipation that
is linear in the velocity gradient.  Landshoff apparently understood that
the overshoot was a result of numerical (truncation) error, but still
attempted to justify the linear viscosity in a more theoretical calculation.
His calculation included integrating a characteristic through the shock,
which effectively implies that entropy is conserved through the shock.
Ironically, this is the same error made by Stokes more than 100 years
earlier, only to be corrected in Rankine's 1870 paper \cite{Salas}. That
entropy is not conserved in a shock but must increase was explicitly
noted in Richtmyer's report \cite{LA671}.  Landshoff also makes 
several errors in his truncation analysis of the discrete equations.

Despite these criticisms, Landshoff's numerical simulations are more
than adequate to justify the {\em linear plus quadratic Q} formulation. 
It is unfortunate that Landshoff's report was not submitted for
publication in the open literature where a competent review might
have led to a more correct exposition.

In a curious historical side note, Harwood Kolsky wrote an internal
Los Alamos report \cite{Kolsky} just a few months before Landshoff.
That report describes one of the first two--dimensional hydrodynamic
simulations. Kolsky discusses in
detail his artificial viscosity, which employs only the Richtmyer--von 
Neumann quadratic term.  He explicitly notes that entropy must be
created in a shock and goes on to suggest the magnitude of the
artificial viscosity is a good indicator of the location and movement
of a shock on the mesh.

\vspace{.12cm}

\noindent
{\bf William Noh}: Bill Noh spent his career at Lawrence Livermore Laboratory,
making fundamental contributions to CFD. These included 
the first method to represent multimaterial cells (SLIC) and an early effort to combine 
the Lagrangian and the Eulerian methodologies (the CEL code).  However Bill's 
lifelong passion was the study of artificial viscosity and its potential pitfalls.

Noh is most well--known for a test problem he devised to illustrate a pitfall
termed wall--heating \cite{Noh,Rider}.  His self--similar problem is designed to 
focus on the excess heating that occurs when a shock is reflected off a wall
in plane symmetry or from a point of convergence in cylindrical or spherical
symmetry. The Noh problem is widely used for code verification \cite{Ramsey,Vel}.

Noh's 1978 paper \cite{Noh} is widely cited, but most of those citations are concerned
with the test problem, missing what we believe is Noh's main point.  In his own
words, 
\vspace{.12cm}

\begin{addmargin}[1em]{2em}
``We want to demonstrate that the wall heating {\em Q} error is unavoidable and 
is already an error in the solution of the differential equations with {\em Q}."
\cite{Noh}
\end{addmargin}
\vspace{.12cm}

\noindent
Here, Noh is saying that the source of the wall heating lies in the {\em model}
itself rather than in the process of discretization.
Noh also proposed a solution to wall--heating, advocating the additional dissipation
associated with an artificial heat conduction and showing it effectively mitigated
the excess heating.  However, artificial heat conduction is not widely used today.
We will expand on this point in section~\ref{discuss}.

In \cite{Noh}, Noh also describes a separate problem when one wants to
propagate a shock on a nonuniform mesh.  Conservation is not a sufficient
condition to ensure the Rankine--Hugoniot jump conditions; it is also
necessary that the shock be steady.  The coefficients of artificial viscosity
depend on the mesh spacing rather than physical length scale.  Most
implementations of artificial viscosity are designed to maintain shock
width of 3--4 cells.  If the mesh spacing is not constant, then the
shock will not be steady.  For more discussion on this problem, 
see \cite{GradScale}.
\vspace{.12cm}

\noindent
{\bf Equations}: In modern notation, the Richtmyer--von Neumann artificial
viscosity is written:
\be \label{eq: Q}
q_{RvN} =   \rho  \, c_Q \, \, \Delta x^2 \, \left(\frac{du}{dx}\right)^2
\ee
Here, $\rho$ is fluid density, $\Delta x$ is the width of the computational cell,
and $u$ is the material velocity.  $c_q$ is a dimensionless constant that is 
set by the user; it is usually chosen $c_q \sim 2$ which serves to spread 
the shock over 3 or 4 cells. For implementation, the artificial viscosity $q$ is
added to the physical pressure in the momentum and energy equations.

The linear plus quadratic viscosity suggested by Landshoff has the form
\be \label{eq: LQ}
q_{LQ} =  \rho \left[
c_Q \,  \, \Delta x^2 \,  \left(\frac{du}{dx}\right)^2  - 
c_L  c_o \Delta x {\frac{du}{dx}} \right]
\ee
Here, $c_L$ is another dimensionless constant set by the user.  $c_o$ is the
fluid sound speed ahead of the shock. The linear term is sometimes
only turned on when the flow is compressive \cite{Caramana}, i.e., 
when $\frac{du}{dx} < 0$ \vspace{.12 cm}

\noindent
{\bf Summary}: Flows with shocks represent many of the earliest applications of 
electronic computers driven by the mission needs of Los Alamos and Livermore
Laboratories.  The lack of resolution of the dissipative length scales of physical
viscosity and heat conduction led to the need for artificial dissipation to avoid 
unphysical oscillations in the solution, i.e., artificial viscosity. 
Even today, it is not practical to fully
resolve shock structure in multidimensional problems of engineering interest.
However, given the utility and the effectiveness of adding an artificial viscosity,
one might ask whether it is necessary or even desirable to fully resolve
shocks, and - if so - what equations should be discretized?  It is well known 
that the Navier--Stokes equations do not accurately predict the details of shock 
structure as can now be measured in experiments, nor is there an accepted 
alternative at the level of continuum theory \cite{Young}.

\begin{remark}
We note that there is no ``correct" shape for the numerical shock profile.
There is a sustained desire among code users to calculate the narrowest
possible shocks that are free of unphysical oscillations. However, these
are competing requirements and in the end, shock profile is determined
by the user's choice of $c_Q$ and $c_L$.
\end{remark}

\section{Nonoscillatory Eulerian Simulation} \label{euler}
The story of nonoscillatory differencing begins with the Barrier Theorem
proved in Godunov's 1954 Ph.D. thesis.   

\begin{addmargin}[1em]{2em}
``Linear numerical schemes for solving partial differential equations having the property 
of not generating new extrema (monotonicity preserving schemes), can be at most 
first--order accurate."  \cite{Godunov}
\end{addmargin}
\vspace{.12cm}

Here, linear means that the numerical discretizations are the same at every grid point,
independent of the flow conditions. Accuracy refers to the dependence of discretization
error on $\Delta x$; error in a first--order scheme is directly proportional to $\Delta x$.
Preserving monotonicity and high--order (or at least second--order) accuracy are both 
desirable properties. Donor cell approximations, also termed upwind differencing 
schemes, take into account the direction of fluid flow, but are only first--order accurate 
and are too diffusive for most applications. For the next 18 years after Godunov, 
Eulerian methods chose accuracy while allowing unphysical oscillations. 

The strategy to bypass the Barrier Theorem was first recognized by Jay Boris and is
surprisingly simple, namely to {\em give up linearity.} In the flux-corrected transport (FCT) 
algorithm \cite{fct}  Boris and Book mix first and second--order methods locally so
as to preserve monotonicity by construction.  FCT uses flow dependent coefficients to 
approximate the advective terms and so is not linear in the sense defined above. It is also 
not quite second-order accurate by standard measures, but can be made very high--order 
in smooth regions of the flow. A fuller history of FCT can be found in \cite{Boris40}.  
Jay Boris is truly the first hero of nonoscillatory methods.

The strategy for avoiding Godunov’s barrier was quickly accepted by the computational
fluid dynamics community and many new methods based on flux limiting evolved.
Those are generally referred to as nonoscillatory finite volume (NFV) methods.
Some elements of the similarity of these many schemes are described in 
\cite{Sweby}.
All NFV schemes are nonlinearly stable under appropriate time step 
restrictions\footnote{nonlinear stability cannot be evaluated using von Neumann's linearized
analysis and must be addressed by more general methods;  see e.g., \cite{Straughan}.},
enforce exact conservation of mass, momentum and energy, and do not allow unphysical
oscillations. NFV schemes are free of parameters so being more automatic and perhaps 
less flexible than explicit artificial viscosity schemes.  Most NFV schemes are built on
the concept of preserving monotonicity and are generally easy to implement in a spatially 
split format on a structured grid.

\begin{remark}
Monotonicity is a mathematical property of a solution and is a one--dimensional
idea.  One expects enforcement of monotonicity is related to the second law 
\cite{MM}; however, this is open to question.  We note the well--known result
of Morduchow and Libby \cite{ML49} that the velocity increases monotonically
in Becker's solution of the shock profile, but the thermodynamic entropy
has a peak inside the profile. This counter--intuitive result is due to an entropy
flux associated with heat conduction, and has been shown to be consistent with
the Clausius--Duhem inequality, which is a local expression of the second--law
\cite{Thompson}.
\end{remark}

\noindent
{\bf Summary:}  The NFV algorithms appear to offer a second, distinct
methodology for controlling unphysical oscillations. Where the artificial viscosity
methods alter the model equations, the NFV methods work more directly by
modifying the solutions.  In the next section, we will describe current research
in which flux--limiting is used to replace the artificial dissipation in Lagrangian 
codes. 

However, the connection between artificial viscosity and NFV methods is
closer than one might expect.  Modified equation analysis (MEA) \cite{MEA} is
a technique to construct a PDE that more closely represents the discretized
equations solved within the code.  A consistent MEA would consist of the
Euler equations plus additional {\em truncation} terms depending on the
discretization parameters $\Delta x$ (the computational cell size) and
$\Delta t$ (the computational timestep).  In \cite{rationale} it was shown
that the NFV method MPDATA \cite{MPDATA} contains quadratic
artificial viscosity as a truncation term.  In chapter~5 of \cite{book}, the
analysis is extended to several additional NFV schemes with the
same result.  \vspace{.12cm}

\begin{remark}
Truncation terms are not necessarily truncation error.
\end{remark}

\vspace{.12cm}

\section{New Ideas in Artificial Viscosity} \label{ideas}
Despite 70 years of effort and the inarguable success of the concept of
artificial dissipation, several practical issues continue to occupy the
attention of code developers with respect to artificial viscosity.  Two of
these are:
\begin{enumerate}
\item
How to choose the optimal values of $c_L$ and $c_Q$ for a particular
simulation.  Those parameters are typically user input and their
selection often determined by trial and error.

\item
How to formulate the artificial viscosity for multidimensional problems
to best preserve the physical flow symmetries and to avoid mesh
dependence.

\end{enumerate}
\vspace{.12cm}

\noindent
{\bf Choosing the viscosity parameters}: The more sensitive issue here
concerns the choice of the linear coefficient $c_L$. As pointed out by
Landshoff, the quadratic viscosity is not large enough to control the
overshoot and postshock oscillations where the velocity gradient is
small.  However, use of the linear artificial viscosity is effectively 
adding a first--order error (i.e. $\Or(\Delta x)$) which can affect
the smooth area of the flow, away from any shocks and where no
viscosity is required \cite{Morgan1}.

An obvious solution lies in the NFV flux--limiting algorithms, which enforce
monotonicity by adding a flow--dependent amount of donor cell, i.e.,
first--order dissipation; effectively these methods methods introduce a 
variable coefficient of linear viscosity automatically. Kuropatenko made
an early attempt to predict the ``optimal" values of $c_L$ and $c_Q$ from 
the solutions of the Riemann problem for a perfect gas \cite{Wilkins}.
The principal result is shown in eq.~(11) of \cite{Wilkins} (reproduced below
as eq~(\ref{eq: Kuro})), an equation that 
predicts the jump in pressure across a shock as a function of the jump
in velocity; this equation goes back to Hugoniot \cite{Johnson},
\be \label{eq: Kuro}
P_1 = P_o + \Gamma \rho_o\left(\Delta U\right)^2 +
\rho_o |{\Delta U}| \left[\left(\Gamma \Delta U \right) ^2+a_o^2     \right]^{\half}
\ee
where $\Gamma = \frac{\gamma +1}{4}$ and $a_o$ is the sound speed.
However, Kuropatenko's interpretation of this equation is erroneous; if
$\Delta U \approx \frac{du}{dx}\Delta x$, then 
$\Delta P \equiv P_1-P_o \approx \frac{dP}{dx}\Delta x$; i.e., this is 
an equation for the pressure gradient, not the pressure.  Further, the
prediction for $c_L$ is not constant, but depends on the shock strength,
e.g., Mach number, which as a practical matter is not known within the
code. It is recognized that the Kuropatenko form overly damps weak
shocks \cite{JasonA}.

Kuropatenko's analysis has led to numerous attempts to adapt
the flux--limiting concept to a staggered mesh Lagrangian code. The earliest
experiment appeared in an unpublished Livermore Laboratory report by  
Christiansen \cite{Randy} in 1990. The Christensen limiter compares the 
velocity divergence in a cell (which is a scalar) with the divergences in the 
neighboring cells. This limiter functions as a shock detector that
identifies cells where limiting should be applied; see section~\ref{heroes}
of this paper.

Subsequent work \cite{Caramana,Loubere,Morgan2} has attempted to improve on
Christiansen's strategy by localizing the flux-limiting within the computational cell 
and by providing a more multidimensional estimate of the direction of the shock.  
In \cite{Morgan2}, a subcell decomposition \cite{Caramana} is employed to integrate 
the momentum equation by implementing Riemann solutions at the subcell boundaries. 
Multidimensionality is introduced by rotating the calculation into the estimated direction 
of the shock. More details details can be found in \cite{Morgan2}


A different and complementary idea to choose the viscous coefficients was
introduced in \cite{Reisner}. In the so-called C--method, an additional 
reaction--diffusion equation and variable is introduced.  The new variable
$C(x,t)$ is used to \vspace{.12cm}
\begin{addmargin}[1em]{2em}
``determine the location, localization, and strength of the artificial viscosity. Near shock 
discontinuities, $C(x,t)$ is large and localized, and transitions smoothly in space--time 
to zero away from discontinuities."  \cite{Reisner}
\end{addmargin}
\vspace{.12cm}

\noindent
 The C--equation is postulated rather than derived and
oddly does not contain an advective term.  We note that the C--methodology
includes an artificial heat conduction (as suggested by Noh) whose coefficient 
is similarly limited by $C(x,t)$. The overall C methodology does not require 
linear artificial viscosity nor linear artificial heat conduction.


Still another approach to generating spatially dependent viscosities is proposed
in \cite{JasonA}. Here, a reference set of {\em optimal} values for the coefficients
$c_L$ and $c_Q$  is pre--computed as a function of shock strength. Then for an 
arbitrary flow, one first estimates the shock strength and then interpolates the 
appropriate coefficients from the reference set. The reference set is
constructed based on modifying the coefficients of the previously mentioned 
Kuroptenko eq.~(\ref{eq: Kuro}).  As noted at the end of section~\ref{heroes} 
there is no correct profile for the numerical shock and {\em in principle} each 
user can weight the relative importance of a steep shock versus the lack of
overshoot based on the purpose of the individual calculation.  There is
a detailed discussion of the construction of the reference set in \cite{JasonA}.

\vspace{.12cm}

\noindent
{\bf Artificial tensor viscosity:} The early hydro codes were written in 1D, limited
by the capabilities of the available computers.  By the early 1960s, however,
2D codes were becoming common \cite{Alder}. Multidimensional Lagrangian
codes are prone to tangling the mesh, and it was quickly realized
that a tensor artificial
viscosity was more robust than a simpler scalar artificial pressure. There are
two principal formulations, edge viscosity and cell-centered viscosity. 
One can think of the edge viscosity of Schulz
\cite{Alder} as nonlinear springs connecting the cell vertices. The cell--centered 
viscosity of Wilkins \cite{Wilkins} is proportional to the velocity gradient within the cell.
These are described and compared in \cite{aspect} including the
important modification to the edge viscosity made by Barton.

Detailed comparisons between the edge and cell--centered viscosities are presented 
in \cite{Campbell}.  The edge viscosities exhibit large mesh sensitivity when the
symmetries of the mesh and of the flow are discordant. In particular, the authors
of \cite{Campbell} suggest that the edge viscosity is not a proper discretization
of a continuum tensor and lacks the invariance properties of a true tensor.
This opens the question about what constitutes a proper discrete calculus for the 
staggered mesh. Here we present a short diversion to another relevant idea in CFD, 
namely {\em mimetic differencing} \cite{reading}.

The general idea of mimetic differencing is to identify and embed certain
properties of the continuum model into the discrete equations. An early
example is the development of finite volume methods by Lax \cite{Lax}.
Whereas finite difference methods provide conservation, e.g., of 
energy, to the level of truncation error, the finite volume methods
ensure conservation to the level of roundoff error.  The underlying idea
is that of a detailed balance in which the {\em flux} of energy out
of a cell is exactly the flux into its neighbor.  We note that detailed
balance does not ensure the flux of energy is accurate, only that
the energy is conserved.

In the case of a discrete calculus, mimetic ideas were first used to ensure the
discrete gradient operator is adjoint to the discrete divergence operator.
This construction was described in another unpublished Los Alamos report
\cite{compatible} and was termed {\em compatible differencing} at the time.
A simultaneous and independent presentation was published in the Soviet
Union and was termed {\em support operator theory} \cite{Shashkov}.

The mimetic approach to defining the discrete differential operators is
based on preserving integral identities.  For example, for any scalar field
$p$ and vector field $\vec v$, the continuous divergence operator $\rm div$ 
and the continuous gradient operator $\rm grad$ are related by 
\be
\int_{V(t)} p\, \rm div(\vec v) \,dV + \int_{V(t)} (\rm grad \,p) \cdot \vec v \, dV = 0
\ee
Here, $V$ is a time--dependent finite volume. 
In the mimetic formulation, we replace the integrals 
by summation; in the particular case of the staggered mesh, the first integral
is replaced by a sum over cells and the second by a sum over vertices. Next,
we assume one of the discrete operators is known, and identify its compatible
adjoint term by term.  Typically, the discrete divergence is chosen as the known 
operator, since the volume of a computational cell is established by Pappus 
theorem and the divergence of the velocity is related to the time rate of change 
of volume.  Details of that straightforward calculation can be found in
\cite{reading}.

For the tensor artificial viscosity, one needs both the gradient of a vector and the
divergence of a tensor evaluated on a nonuniform computational mesh.  Those operators
are formulated in \cite{Campbell}.  They are complicated; however the improvement
in preserving symmetry is also substantial as can be seen by comparing figures~8 
and 10 of that paper.

Another important idea that was unveiled in \cite{aspect} concerned the {\em model} form 
of the tensor viscosity. There it is shown that the artificial tensor viscosity
$Q_{ij}$ should be taken to be directly proportional to the velocity gradient 
\be \label{eq: Qij}
 Q_{ij} \sim \frac{\partial u_i}{\partial x_j} \,.\ee
rather than to the symmetrized gradient.  This choice ensures that there is no {\em
mode conversion}; i.e., that in a shear flow in which all velocities are parallel,
the artificial viscosity will not generate velocity components perpendicular
to the flow. The importance of suppressing mode conversion was illustrated in
the Saltzman test problem in which a one-dimensional shock propagates
through a two-dimensional mesh.  The same choice of using eq.~(\ref{eq: Qij}) is
made in \cite{Campbell}. We will discuss this choice further in section~\ref{discuss}.

\section{The Physics of Finite Volumes} \label{FV}
The Finite Volume Method (FVM) \cite{Leveque} is a popular methodology in CFD. 
In FVM, the discrete variables represent volume averages over the 
computational cells rather than values at a particular point within the cell. FVM
methods are exactly conservative of mass, momentum and total energy at 
the level of round off error as contrasted with finite difference methods, which are
conservative only at the level of the truncation error.  FVM originated at Los
Alamos in an unpublished report by Peter Lax \cite{Lax}.  
\vspace{.12cm}

\noindent
In \cite{rationale} Margolin and Rider posed the following question: 

\begin{addmargin}[1em]{2em}
``If every point of a finite volume of fluid is governed by the Navier–Stokes equations, 
what equations describe the evolution of the volume averages of the state variables?
\end{addmargin}
\vspace{.12cm}

\noindent
Answering this question led to the finite scale theory.  Although  \cite{rationale}
is ostensibly concerned with rationalizing the methodology of implicit large
eddy simulation, its two principal results are more general. These are:
\vspace{.12cm}

\noindent
(1) A general derivation of the evolution equations, termed the Finite Scale
equations (FSE) for the volume--averaged
variables of compressible fluid flow;\\
(2) An analysis showing that a particular NFV, i.e., MPDATA, solves the
FSE in the sense of modified equation analysis \cite{MEA}.
\vspace{.12cm}

\noindent
The FSE have the form of augmented Navier--Stokes equations.  New terms 
appear proportional to the constant $A$, which has the dimensions of length squared.
In one spatial dimension when the equation of state is a perfect ($\gamma$-law) 
gas, the FSE are written: 
\be \label{eq: rho}
\frac{\partial \bar \rho}{\partial t} + \frac{\partial \bar \rho \tilde u}{\partial x} = 0
\ee
\be \label{eq: mom}
\frac{\partial \bar \rho \tilde u}{\partial t} + \frac{\partial}{\partial x}\left(\bar \rho (\tilde u)^2 + P\right)
=0
\ee
\be \label{eq: TE}
\frac{\partial}{\partial t}\left(\tilde \I+\half \bar \rho (\tilde u)^2\right) +
\frac{\partial}{\partial x}\left(\tilde \I \tilde u + \half \bar \rho (\tilde u)^3+P\tilde u+Q \right) = 0
\ee
\be \label{eq: P}
P = (\gamma -1)\bar \rho \tilde \I - \mu \, \frac{\partial \tilde u}{\partial x} 
+  A\bar  \rho \left(\frac{\partial \tilde u}{\partial x}\right)^2
\ee
\be \label{eq: QH}
Q =  -\kappa \frac{\partial \tilde \I}{\partial x} +
\gamma A \bar \rho \left(\frac{\partial \tilde u}{\partial x}\right)\left(\frac{\partial \tilde \I}{\partial x}\right) 
\ee
Here, the overbar signifies a volume--averaged quantity, 
\be \label{eq: aver}
\bar \rho(x) \equiv \frac{1}{\Delta x}  \int^{x+\Delta x/2}_{x-\Delta x/2} \,\rho(x') \,dx'  \ee
The tilde indicates a Favre-averaged quantity.  For example,
\be \label{eq: Favre}
 { \tilde \I}(x) \equiv \frac{1}{\bar \rho(x)\Delta x}
 \int^{x+\Delta x/2}_{x-\Delta x/2} \,\rho(x')\,\I(x') \,dx'  \ee
 where $\I$ is specific internal energy.  Also, $\mu$ and $\kappa$ are the coefficients
 of physical viscosity and heat conduction. These coefficients represent processes
on the length scale of the molecular mean free path $\lambda$ and the
associated terms are negligible when $\Delta x \gg \lambda$.
Note that FSE is a PDE theory of volume--averaged variables rather than an
integral theory of differential variables.
 
The constant $A \equiv \LL$ where $\Delta x$ is the length of the (1D) 
finite volume.  In the case of numerical simulation, $\Delta x$ is the
size of a computational cell.  However, the FSE are analytic equations
and $\Delta x$ may be chosen arbitrarily to resolve the process or
perhaps the measuring instrument of interest.  With reference to the
Heisenberg quote in the introduction, we refer to $\Delta x$ as
{\em the observer}, so quantifying the ``we" in the quote.

\begin{remark}
In principle the derivation of the FSE leads to an infinite series of terms
in even powers of $\Delta x$.  Equations~(\ref{eq: rho})--(\ref{eq: QH})
are a truncation of the equations at $\Or(\Delta x^2)$.
\end{remark}

Note that a quadratic
viscosity term analogous to the Richtmyer--von Neumann term appears in 
eq.~(\ref{eq: mom}) and a nonlinear heat conduction term analogous to that 
proposed by Noh appears in eq.~(\ref{eq: TE}) . By analogous, we mean that 
in addition to the functional form, the coefficient $A$ of the new terms is {\em inviscid},
i.e., is not property of the fluid.  However, the analogy does not extend to
include a linear inviscid viscosity; such a term is generally ruled out by the 
isotropy of space. 

The derivation of the FSE begins at a small  length scale $\Delta x_{\epsilon}$  
such that all gradients 
of the flow variables are well--resolved.  At such a scale, it is assumed that the 
Navier--Stokes equations are germane and that the dependent field variables can 
be expanded in a convergent Taylor series.  The steps of the calculation can be 
summarized as follows.
\begin{enumerate}
\item
Insert the Taylor series expansions into the Navier--Stokes equations and integrate
the equations over\\
$x \in [x-\Delta x_{\epsilon}/2,x+\Delta x_{\epsilon}/2]$. Carefully commute the
integration and the differential operators, then truncate the results to 
$\Or(\Delta x_{\epsilon}^2)$. This leads to the FSE at scale is $\Delta x_{\epsilon}$.

\item
Assume the form of the equations holds for arbitrary $\Delta x$ and derive the equation
for $2\Delta x$.  Show that the form is unchanged under that renormalization,
proving {\em form invariance} of the FSE.  

\end{enumerate}
Details of both steps can be found in \cite{rationale,Plesko}.  Here, we discuss four further 
salient features of the calculation.  First, there is a theorem of closure.  We note that
multiplication and integration do not commute,
i.e., ${\tilde u}^2 \ne \widetilde {u^2}$.  It is that lack of commutativity due to the 
{\em nonlinearity} of advection that leads to the inviscid dissipative terms of the FSE. In general, 
the calculation leads to a closure theorem for finite scale variables: 
\vspace{.12cm}

\noindent
{\bf Closure theorem}: For any continuum fields $A$ and $B$ that are sufficiently smooth
at small scales, 
\be \label{closure}
\overline {BC} = \overline B \, \overline C + 
\frac{1}{3}\LL \, \frac{\partial \overline B}{\partial x}\,\frac{\partial \overline C}{\partial x}
+ \Or(\Delta x^4)
\ee
A similar theorem applies to Favre averaging.

Second, we observe that at every length scale $\Delta x$, there are 3 contributions to
the momentum flux; these are the advective flux, the viscous diffusive flux and the
inviscid diffusive flux.  As one considers progressively larger averaging scales, the
inviscid flux grows while the advective flux decreases.  This happens so as to ensure
the total momentum flux is conserved.  There is an analogous result for the energy
flux.  Details can be found in \cite{Plesko}.

Third, the form of the FSE has the interesting interpretation that many features of the 
large scales of the flow do not depend on the details of the small scale processes.  
For example, in a shock, the 
velocity and jump conditions are independent of the viscosity; only the detailed shape 
of a shock depends on the fluid viscosity \cite{Taylor}. In turbulent flow, the energy
dissipated is determined by the large scales of the flow; viscosity only determines
the size of the smallest eddies at which dissipation occurs; see the discussion
of Kolmogorov's $4/5^{th}$ theorem in \cite{Frisch}.  In the theory of
nonlinear PDEs, this property indicates the existence of an inertial manifold
\cite{Foias}.  One might say that small-scale variables are only needed to
predict the small-scale features of a flow, and it is the large-scale dynamics that 
determine how the overall flow behaves. Or, in other words, when describing the 
large-scale evolution of a flow, the details often do not matter.

Fourth, we comment further on the lack of a linear inviscid term in the FSE.  In
 \cite{TWFS} analytic solutions of equations with both linear and quadratic artificial
 viscosity were compared to solutions with only quadratic viscosity.  In the case
 with linear artificial viscosity, the solutions were not compact, indicating the
 unphysical property that information can travel with ``infinite" speed. However,
solutions of equations with only quadratic viscosity have compact support.
Although this is a positive feature for FSE, it indicates a potentially larger
issue, namely that the derivation of the FSE begins at the smallest scales 
with Navier--Stokes and so inherits some of the issues of that continuum
model.

\section{A Physical Interpretation} \label{flux}
The presence of inviscid fluxes in the mathematical derivation is suggestive,
but a physical interpretation of those fluxes would be even more compelling.
Such an interpretation was first offered in \cite{royal}.  We reiterate that
the FSE are not integral equations for continuous fields, they are PDEs for
volume--averaged fields. \vspace{.12cm}

\noindent
{\bf Inviscid fluxes}: 
Consider integrating a 3D extension of eq.~(\ref{eq: rho}) over a volume $\Omega$. 
Here we have replaced the 1D partial derivative with the 3D divergence.  Invoking
Reynolds transport theory and using the divergence theorem, 
\be
\frac{d}{dt}\left(\int_{\Omega} \overline \rho \,dV    \right) = \int_{\partial \Omega} 
\overline \rho\, (\tilde u - w)\cdot\ n \,  dS              
\ee
where $w$ is the velocity of the surface and $n$ is a unit normal to the surface.
The term on the left is the time rate of change of the mass of $\Omega$.
When we choose a {\em Lagrangian volume} whose surface moves with
the local fluid tilde velocity, the term on the right vanishes indicating that the
mass of $\Omega$ is constant. From the kinetic point of view, this does
not mean that every fluid particle initially in the cell remains in the cell
for the duration of a calculation.  Rather, we infer the weaker statement
that the net flux of mass integrated around the total surface of the volume
vanishes. 

Now we note that the vanishing of the mass flux does not imply the
vanishing of the momentum flux nor of the energy flux.  In \cite{royal}, 
estimates support the interpretation that these {\em exchange} fluxes
are the source of the inviscid fluxes in the FSE.
This interpretation gives substance to Bill Noh's assertion that wall
heating results from an inconsistent model rather than its numerical
implementation.  A system with artificial viscosity in the momentum
equation, but without artificial heat conduction in the energy
equation is simultaneously open and insulated, which is a
thermodynamic inconsistency. \vspace{.12cm}

\noindent
{\bf Discrete thermodynamics}:
As noted in the previous section, the inviscid fluxes originate in the
nonlinearity of advection.  There is another important nonlinearity in
the FSE for a perfect gas, namely in the equation 
of state.
\be 
\bar p = (\gamma-1)\overline {(\rho\, \I)}
\ee
Mathematically, the averaged term on the RHS can be resolved by
the closure theorem.  However, in the physics there are deeper
issues that can only be resolved within the perspective of kinetic
theory.

In \cite{Hunter}, the {\em nonequilibrium} of finite volumes was discussed.
Here, nonequilibrium implies the presence of macroscopic gradients,
e.g., of velocity or internal energy or density. There it is demonstrated
that, in the context of finite volumes, temperature and internal energy 
are not equivalent. Temperature is a purely equilibrium concept while 
the internal energy also contains the {\em unresolved} kinetic energy. 
The latter is an independent partition of the total energy, which is fed 
by inviscid fluxes and which decays into internal energy on flow-dependent
time scales. We note the similarity of this result to the models of
turbulent kinetic energy (TKE) commonly used in atmospheric flows 
\cite{Pope}. 

Finally,  the existence of significant unresolved kinetic
energy implies that the assumption of local thermodynamic equilibrium
(LTE) is not justifiable \cite{MRJ}.  LTE is a basic assumption in the 
derivation of the Navier--Stokes equation from the Boltzmann equation 
via the Chapman--Enskog approximation; its failure is another
reason to seek a derivation of the finite scale equations beginning
at the level of gas kinetic theory. \vspace{.12cm} 

\section{Discussion} \label{discuss}
The finite scale theory provides an understanding and theoretical basis for 
traditional viscosity as well as for many of the ideas introduced more
intuitively in section~\ref{ideas}.  Here we expand on several topics:
\vspace{.12cm}

\noindent
{\bf Formulation}: The quadratic form of viscosity a la Richtmyer and von Neumann, shown in
eq.~(\ref{eq: Q}), is predicted by the FSE in eq.~(\ref{eq: P}). Its inviscid
character reflects its origin in the nonlinearity of advection rather than in
dissipative processes of collisions, i.e., the inviscid coefficient is geometric and
independent of the molecular mean free path.  The quadratic viscosity
should be viewed as a flux of momentum rather than as an addition
to the pressure.

The derivation of the FSE precludes a physical basis for Landshoff's linear
artificial viscosity.  Furthermore analytic solutions of the FSE show that the
shock profile is monotonically increasing in velocity, density, internal
energy, etc. \cite{TWFS}  This indicates that truncation error is the source 
of overshoots and unphysical oscillations. Careful truncation analysis 
in \cite{structure} shows the error term causing the oscillations is $\Or(\Delta x^2)$; 
the linear artificial viscosity is a heavy--handed mitigation that unnecessarily alters
the smooth areas of the flow.

In eq.~(\ref{eq: QH}) the FSE also predicts a quadratic artificial heat conduction, 
similar to the form suggested by Noh in eq.~(2.4) of \cite{Noh}. However, the
linear analog of artificial heat conduction is also precluded in the FSE. It is 
interesting to note that the inviscid Prandtl number, i.e., the ratio of the inviscid
momentum diffusivity to the inviscid internal energy diffusivity is predicted
to be ${1}/{\gamma}$.  This is exactly the value of the viscous Prandtl
number assumed by Becker \cite{Becker} that enables his analytic
solution. See footnote~\#1 in \cite{MRJ} for more details.
\vspace{.12cm}

\noindent
{\bf Bi--velocity hydrodynamics}: The inviscid fluxes of section~\ref{flux} are transported 
with a velocity relative to $\tilde u$. Indeed, there are two velocities that appear
in the FSE, the average material velocity $\bar u$ and the Favre material
velocity $\tilde u$. A relation between these is derived in \cite{royal} using
the closure theorem:
\be \label{eq: bivel}
 \tilde u = \bar u + \frac{1}{3} \LL \frac{\bar u_x\bar \rho_x}{\bar \rho}
\,.\ee
where $\bar \rho_x = \frac{\partial \bar \rho}{\partial x}$, etc.  The FSE
are written in terms of $\tilde u$, which is the momentum velocity.
$\bar u$ is the advective velocity.  It is the relative velocity $(\bar u - \tilde u)$
that transports the inviscid fluxes.

The subject of {\em bi-velocity hydrodynamics} was first introduced by
Howard Brenner in \cite{Brenner} and its consequences were explored in 
the context of continuum dynamics.  

\begin{addmargin}[1em]{2em}
``Acceptance of the Navier–Stokes–Fourier equations as the fundamental equations 
of continuum fluid mechanics for liquids and gases is noted to be inseparably linked
to Euler’s implicit, but unproved, hypothesis that but a single-velocity field is required 
to characterize the four physically different velocities appearing in 
the mass, momentum, and energy equations."  \cite{BrennerBi}
\end{addmargin}
\vspace{.12cm}

\noindent
Brenner proposes a constitutive relation that in our notation in 1D is:
\be
\bar u = \tilde u + \alpha_v \frac{1}{\bar \rho}\frac{\partial \bar \rho}{\partial x}
\ee
where $\alpha_v$ is a constant termed the ``volume diffusivity".
This is consistent with the FSE relation~(\ref{eq: bivel}) if instead of a
constant we take.
\[ \alpha_v = - \frac{1}{3} \LL \tilde u_x \,.  \]
Brenner terms the velocity difference $(\bar u - \tilde u)$ the {\em diffusive 
volume flux density}.
\vspace{.12cm}

\noindent
{\bf Angular momentum}: In simulations in multiple spatial dimensions, concern
for accuracy must be supplemented by a concomitant concern for the
preservation of the geometric flow symmetries. Unlike the exact conservation
of mass, momentum and energy in the FVM, angular
momentum is only approximately conserved.  The {\em advection} 
interpretation of artificial viscosity in multiple dimensions justifies
the tensor form of eq.~(\ref{eq: Qij}).

Consider again the situation of section~\ref{flux}.
In the process in which there is no net exchange of mass, one cannot
expect that all particle velocities are normal to the surface.  In general,
the particle exchange will generate a tangential force on the boundaries
between cells. However, a cell is not an isolated rigid body and the net
deformation of all the cells must be compatible, i.e., the displacements of
all the cells must be continuous and single--valued.  Solving the
compatibility equations would be a nonlocal problem to be avoided 
in an explicit hydrocode.

In \cite{aspect} a different approach is followed.  It is suggested that
the edge viscosity leads to a better approximation of a compatible solution
since the forces are shared by both cells adjacent to the boundary.
Then a cell--centered (true) tensor viscosity is derived that most closely
reproduces the forces of the edge viscosity.  That tensor turns out to
have the form of eq.~(\ref{eq: Qij}), a nonsymmetric tensor. It is
perhaps ironic that Schulz recognized that his edge viscosity was
not a symmetric tensor and considered that a weakness.
\section{Conclusions} \label{conclude}
From combustion engines to the collision of gases in giant galaxy clusters, 
shock waves are ubiquitous in our universe.  However, as first pointed out 
by Becker in 1922, the continuum models we use to describe them do not 
accurately predict the experimentally measured shapes or widths of the 
shock profile. The use of electronic computers has further exacerbated the 
situation as the dissipative process of viscosity and heat conduction are 
not resolved in typical engineering calculations.

In this paper we have attempted to summarize the historical development of 
modeling shocks on the computer, celebrating its early pioneers, giving a modern 
perspective of its theoretical essence, as well as offering insights for future 
improvements. The underlying idea of shock capturing, first elucidated by Rudolf 
Peierls \cite{Archer} preceded the advent of the first electronic computer 
\cite{Goldstine,Morgan} and is based on the fundamental results of Rankine
\cite{Rankine}. The importance of dissipation in regularizing the 
theoretical shock structure is attributed to Becker in both \cite{LA671,VNR}.  
The importance of heat conduction in shock structure is also emphasized by Becker.  
This is a key idea in Bill Noh's work \cite{Noh} though no attribution is made.  
Nonetheless, all of these works have underlined the utility and 
necessity of artificial dissipation schemes when modeling shocks.

There is a broad theme woven into this paper, namely the close connection 
between theory and successful numerical algorithms. The flow of ideas from
physics to methods is termed mimetic differencing, described in section~\ref{ideas}.
However, the quadratic viscosity has followed the opposite path, originating in 
numerical methods and finding a realization in the theory of finite scales. That 
the Finite Scale Equations -- a continuum description of the discrete numerical 
equations -- contain new {\em inviscid} dissipative terms with an underlying 
physical interpretation suggests that  the "observer" ($\Delta x$) plays a key 
role in consistently describing physical processes across a range of scales 
\cite{observer}. Further, inviscid dissipation implies that the large scales
``control" the small scales, indicating that in many situations
the details of the dissipative processes don’t matter. This point of view was 
clearly expressed in the landmark paper of von Neumann and Richtmyer 
\cite{VNR}.

Artificial viscosity is one of the oldest and most enduring concepts in computational
fluid dynamics.  In particular, the quadratic dependence of the artificial viscosity
on the velocity gradient is now appearing in many different venues, e.g., 
models \cite{Plesko}, nonoscillatory differencing \cite{rationale}
and theoretical physics \cite{Young,Rankine}. We are especially pleased
to reveal the origin of this almost {\em magical} term in the unpublished work of
Bob Richtmyer \cite{LA671,LA699}.

The continued success of artificial viscosity suggests that other concepts from
numerical methodology may have physical relevance in theoretical fluid
dynamics.  Just as role of the observer reminds one of the emphasis on
measurement in quantum theory, the Courant--Friedrichs--Lewy (CFL) condition 
for numerical stability has a causal flavor limiting the speed with which
information can be propagated across the mesh in explicit simulations. 
Generalizations of monotonicity preservation may be related to the dynamics 
of irreversible processes. 
 
The main point of Becker's paper was to point out the inadequacy of
Navier--Stokes theory in predicting shock structure.  So far, no
continuum model has risen to that challenge.  Perhaps it would be
beneficial to look to successful numerical methodology for further
improvements and a more fundamental insight to shock theory
and perhaps more generally to fluid dynamics.

\begin{acknowledgement}
We gratefully acknowledge many illuminating discussions with W. Rider, 
S. Runnels, P. Smolarkiewicz, and D. E. Vaughan.  We thank
N. Morgan, J. Reisner and M. Shashkov for their constructive comments, 
especially of material in section~\ref{ideas}.
This work was performed under the auspices of the U.S. Department of Energy's NNSA 
by the Los Alamos National Laboratory, is managed by Triad National Security, LLC
for the National Nuclear Security Administration of the U.S. Department of Energy 
under contract 89233218CNA000001.
\end{acknowledgement}

%
 \section*{Conflict of interest}
 The authors declare that they have no conflict of interest.



\newpage
\end{document}